\documentclass[aps,pre,final,twocolumn,showpacs]{revtex4}
\usepackage{psfrag}
\usepackage{graphicx,psfrag,subfigure}
\usepackage{amsmath}
\usepackage{bm}
\usepackage{amsfonts}

\begin{document}

\title{Additive Nonparametric Reconstruction of 
Dynamical Systems from Time Series}

\author{Markus Abel}
\email{markus@stat.physik.uni-potsdam.de}
\author{Karsten Ahnert}
\author{J\"{u}rgen Kurths}
\author{Simon Mandelj}
\affiliation{
Institute of Physics, University of Potsdam,\\
14415 Potsdam, Germany
}
\date{\today}
\begin{abstract}
{We present a nonparametric way to retrieve an additive system of differential equations
  in embedding space from a single time series. These equations can be treated
  with dynamical systems theory and allow for long term predictions. We
  apply our method to  a modified chaotic Chua 
  oscillator in order to demonstrate its potential.}
\end{abstract}

\maketitle
                                   
Casting physical observations into mathematical equations is one of the
fundamental tasks to understand and predict dynamical systems. Basically,
there are two complementary approaches to accomplish this task: theoretically
by convenient considerations and empirically by data analysis.  Both
approaches are essential for modern modeling strategies.  If For many systems,
the dynamics is not directly accessible to theoretical considerations; then an
appropriate data analysis is essential.  This problem is very general, one can
find it in classical fields of physics, e.g. classical mechanics, fluid
dynamics, solid state physics, statistical physics as well as in more
interdisciplinary fields, e.g., physiology, earth sciences, economics or
biological systems.  In this paper the data analysis issue is addressed: we
determine an analytically treatable set of additive equations in embedding
space by the method of nonparametric embedding. This approach is {\it a
priori} parameter-free; but {\it subsequent} parameterization can be helpful
for analytical representation of the involved functions.

Often, the measurement of a complex system does not yield the whole set of
state variables. The missing dynamics can be accessed by the {\it embedding}
technique \cite{Takens-81}. Given the measurement of a subset of variables,
one can infer the missing information by an embedding map, e.g., by using the
time-delayed variables or their derivatives. This has been proven rigorously
for a wide class of systems \cite{Sauer-Yorke-Casdagli-91}. It is, however,
not known how the equations of the dynamical system in embedding space are
structured. In this communication, we propose a technique to find a set of
equations which allows a reproduction of the dynamics in phase space for the class
of {\it additive} systems.

There are several excellent reviews about embedding
\cite{Sauer-Yorke-Casdagli-91,Kantz-Schreiber-97,Abarbanel-97}; therefore, we
only repeat some basic facts.  We consider a system governed by a set of
ordinary differential equations:
\begin{equation}
\label{eq:flow}
\dot{\vec{x}} = {F}(\vec{x})\;,
\end{equation}
where $\vec{x}\in \mathbb {R}^n$, $F: \mathbb{R}^n\mapsto\mathbb{R}^n$. This
set of equations defines a flow, $F_t$, in phase space. We assume that there
exists an attractor $\cal{A}\subset$ $\mathbb{R}^n$ with the box-counting
dimension $d\leq n$. In \cite{Sauer-Yorke-Casdagli-91} it has been shown that
almost every smooth map $\Psi: \mathbb {R}^n \mapsto \mathbb{R}^m$, $m>2d$, is
an embedding, i.e. a smooth diffeomorphism from $\cal{A}$ onto its image
$\Psi(\cal{A})$. The condition $m>2d$ is sufficient, therefore cases
with $d<m<2d$  can occur.

Due to differentiability, the dynamics of $\vec{\xi}(t)=\Psi(\vec{x}(t))$ obeys an
ordinary differential equation in embedding space:
 \begin{equation}
\label{eq:embflow}
\dot{\vec{\xi}} = {\Phi}(\vec{\xi})\;,
\end{equation}
with $\vec{\xi} \in \mathbb{R}^m$, ${\Phi}: \mathbb{R}^m\mapsto\mathbb{R}^m$.  In
this article, we focus on {\it additive} models for the components $\Phi_i$
and show how to retrieve them from data.

One standard way of embedding is the use of the delay-coordinate map
$H(f,\tau)$, with the smooth observation function, $f: \mathbb{R}^n \mapsto
\mathbb{R}$, and $\tau$, the time delay, some real number
\cite{Sauer-Yorke-Casdagli-91}:
\begin{equation}
H(f,\tau) = \left(f,\,f(F_{-\tau}),\, \dots\,,f(F_{-(m-1)\tau})\right)\;.
\end{equation}
As an example, consider the particular case of $f(\vec{x})=x_1$. Identifying
the above embedding map $\Psi$ with $H$,  the coordinates in embedding
space are $\xi_1(t)=f(\vec{x})=x_1(t),\xi_2(t)=f(F_{-\tau}(x))=x_1(t-\tau)$,
etc..

In our analysis, we perform numerical simulations for some model systems to
obtain time series of various variables.  We then discard all but one variable
to embed the dynamical system (\ref{eq:embflow}) using the delay map. To avoid
confusion, we will refer to dynamics from Eq.~(\ref{eq:flow}) as {\it
original}. For the counterpart, Eq.~(\ref{eq:embflow}), to be estimated by
nonparametric regression, we will use the term {\it reconstructed}. If the
embedding map $\Psi$ is concerned, {\it embedded} will be used - the latter
meaning that a time series from the original system is used, i.e. without
knowing the dynamical system (\ref{eq:embflow}).

To find a dynamical system in embedding space, several approaches exist,
e.g., local linear fits and parametric procedures as polynomial fits, radial
basis functions or neural networks (cf. \cite{Kantz-Schreiber-97}).  Local
fitting is a general concept, but the results are neither easy to access
analytically nor to visualize due to the high dimensionality. Polynomial
ansatzes tend to involve too many terms for a clear identification of a
mathematical or physical structure; for neural networks a physical
interpretation is very hard.

Now we describe our procedure in more detail: Considering each temporal
measurement as a realization of the flow, one obtains as a best estimator of
the components of Eq.~(\ref{eq:embflow}) in the least-square sense
\cite{Kantz-Schreiber-97}
\begin{equation}
\label{eq:MD}
\Phi_i = E\left[\dot{\xi_i}\, |\, \xi_1,\dots,\xi_m\right]\;, 
\end{equation}
with $E[\cdot|\cdot]$ the conditional expectation value (CEV) operator. It is
a very hard task to extract analytical models from Eq.~(\ref{eq:MD});
visualization is obviously impossible for $m>2$.

To tackle this problem, we require the rhs of Eq.~(\ref{eq:embflow}) to 
 be an additive model:
\begin{equation}
\label{eq:AddModel}
\Phi_i =  \sum_{j=1}^m \phi_{ij}(\xi_j)\;.
\end{equation}
This is a subset of the class of models considered by Kolmogorov
\cite{KV}: he showed rigorously that it is possible to represent any
continuous function of a set of $m$ variables as a $2m+1$-fold
superposition of $m$ functions of one argument.  Below, we show that
despite the less general formulation it is possible to reconstruct a
chaotic dynamical system.  Our model (\ref{eq:AddModel}) is, however,
still in a wider model class than in parametric methods, because we do
not rely on a given set of basis functions.  After having finally
estimated the components $\phi_{ij}$, we can easily visualize the
functions and try analytical formulae.

It is worth noting the geometrical aspect of our approach:
Eq. (\ref{eq:embflow}) defines a differentiable manifold approximated by the
sum of the functions $\phi_{ij}$, cf. Eq.~(\ref{eq:AddModel}). This is
possible within a certain scatter, which is quantified below by the
correlation. If the manifold is found exactly by the model, the correlation is
1.  Dynamical and topological properties of the original system are mirrored
in embedding space. Long-term predictions of the dynamics are thus possible on
the basis of the obtained model if the correlation is close to 1, which is a
very strong advantage.

The optimal estimate for the $\phi_{ij}$ is calculated by the backfitting
algorithm \cite{Backfit}. It works by alternately applying the CEV operator to
projections of $\Phi_i$ on the coordinates:
$\phi_{ij}(\xi_j)=E[\dot{\xi_i}-\sum_{k\neq j} \phi_{ik}\; |\; \xi_j]$, and is
proven to converge to the global optimum in the least-square
(Eq.~(\ref{eq:chi2})) or correlation (Eq.~(\ref{eq:Cij})) sense.  For the
application to spatio-temporal data analysis, see
\cite{Voss-Buenner-Abel-98,Abel-04}. We calculate the CEV by smoothing
splines, which are optimal for nonparametric regression \cite{Backfit}, due to
their smoothness and differentiability properties. It is important to note
that the parameters used by splines or other estimators are method-inherent
and not prescribed by a preselected model; in this sense the model is
parameter free.

As an overall quality measure, the least-square error can be used
\begin{equation}
\label{eq:chi2}
\chi_i^2 = E\left[ ( \Phi_i - \sum_{k=1}^m \phi_{ik} )^2\right]\;.
\end{equation}
The backfitting method, however, is formulated as optimal in the sense of
correlation, i.e. the natural measure is the correlation coefficient $C_{i0}$
between rhs and lhs in Eq.~(\ref{eq:AddModel}). The correlation coefficient
$C_{ij}$, given in Eq.~(\ref{eq:Cij}), indicates its individual weight for the
model:
\begin{equation}
\label{eq:Cij}
C_{i0}= C\left[ \Phi_i\,;\,\sum_{k=1}^m \phi_{ik} \right]\;\;,\;\;\;
C_{ij}= C\left[ \phi_{ij}\,;\, \Phi_i - \sum_{k\neq j}^m \phi_{ik} \right]\;.
\end{equation}
We will use $C_{i0}$ as a quantitative measure, but not $\chi^2$.  Putting $
\Phi_i=Y$, $\sum_{k=1}^m \phi_{ik}=X$, we give the relation between both
measures: $2\,\cdot\,C_{i0}\,\cdot \sqrt{VAR(X)\cdot VAR(Y)}=VAR(X) + VAR(Y) +
[E(X-Y)]^2 - \chi_i^2$, with $VAR$ the variance.  A correlation close to 1
means the manifold described by Eq.~(\ref{eq:embflow}) is approximated very
well, lower correlations indicate scatter of data points around the
manifold. In the case of experimental data, measurement noise can produce some
additional scatter.

In the following, the procedure is illustrated by the example of a modified
Chua circuit \cite{Madan-93} with a third order nonlinearity. The basic
equations read:
\begin{equation}\label{eq:chua}
\begin{split}
\dot{x_1}= a(m_0 x_1-1/3\, m_1 x_1^3 + x_2)\;,\;\;\\
\dot{x_2}=x_1-x_2+x_3\;,\;\;
\dot{x_3}=-b x_2 \;.
\end{split}
\end{equation}
Written as an additive model (\ref{eq:AddModel}) these equations read
$
\dot{x_1}= f_{1,1}(x_1) + f_{1,2}(x_2)\;,\;
\dot{x_2}= f_{2,1}(x_1)+ f_{2,2}(x_2) +  f_{2,3}(x_3)\;,\;
\dot{x_3} = f_{3,2}(x_2)\;,
$
%
with the linear functions $f_{1,2}$, $f_{2,i}$, $f_{3,2}$  and $f_{1,1}$ 
a third order polynomial.

We integrate the system (\ref{eq:chua}) with $a=18$, $b=33$, $m_0=0.2$,
$m_1=0.01$ numerically by a Runge-Kutta algorithm of 4th order. The time series
of the first component is used for embedding.  Results do not change for other
components.  We first discuss embedding dimension $m=3$. From the time series
$x(t)$, the  points
$\dot{\xi}_i=\dot{x}(t-\tau(i-1)),\xi_1=x(t),\xi_2=x(t-\tau),\xi_3=x(t-2\tau)$ are used. The
derivative is taken directly from the integration, this is more exact than the
estimate by finite differences.  The nonparametric regression yields the
functions $\phi_{ij}$ for the resulting dynamical system (\ref{eq:AddModel}) .

\begin{figure}
  \begin{center} 
 \includegraphics[draft=false,clip=true,angle=0,width=0.4\textwidth]{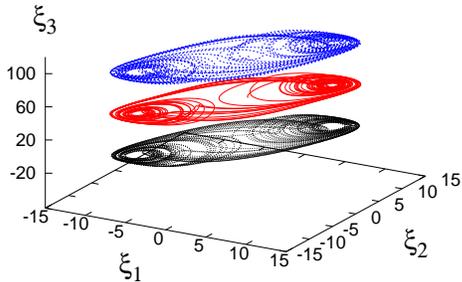}
  \end{center}
\caption{Embedding of the first component of the system (\ref{eq:chua}) with
  delay $\tau=0.2$. The system, embedded with $m=3$ (bottom) is shown together
  with the reconstructed trajectory from the integration of the systems
  obtained by nonparametric regression for embedding dimension 3 (middle) ,
  and 4 (top), respectively. An offset is added to avoid overlap of the
  attractors.}
\label{fig:ChuaEmb}
\end{figure}

\begin{figure}
\begin{center}  
\includegraphics[draft=false,clip=true,angle=0,width=0.4\textwidth]
		      {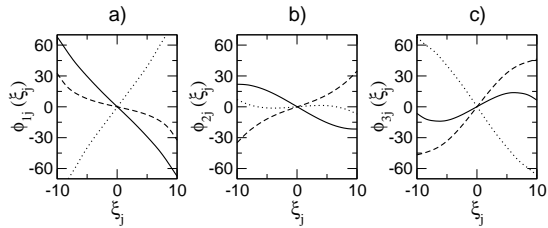}
\end{center}
  \caption{Reconstructed functions $\phi_{ij}$, $i=1,2,3$  in embedding
    space. 
a) $\phi_{1j}$,
b) $\phi_{2j}$,
c) $\phi_{3j}$, for $j=1$ (solid line), $j=2$ (dotted line), and $j=3$ (dashed
    line). All functions are important with $C_{11}=0.999$, $C_{12}=0.99$ ,
    $C_{13}=0.998$ , $C_{21}=0.999$ , $C_{22}=0.991$ , $C_{23}=0.999$ ,
    $C_{31}=0.997$ , $C_{32}=0.999$ , $C_{33}=0.999$. }
\label{fig:ChuaEmbRecFuncs}
\end{figure}

First, we present results for the specific delay, $\tau=0.2$, we study below
the dependence of the results on the delay time.  For $\tau=0.2$, the embedded
and the reconstructed attractor are shown together in Fig.~\ref{fig:ChuaEmb}.
With respect to the data analysis, we want to  quantify
i) the quality of the regression, ii) the importance of the functions
$\phi_{ij}$, and iii) the functions themselves.  

i) \underline{\it The quality of the regression} is given by the correlation
$C_{i0}$, cf. (\ref{eq:Cij}). We find in our case $C_{10}=0.992$,
$C_{20}=0.999$, $C_{30}=0.995$, such that the modeling error is very small.

ii)\underline{\it The importance of functions} is found by the coefficients
 $C_{ij}$, defined in Eq.~(\ref{eq:Cij}) ($i,j=1,2,3$). We find $C_{ij}>0.99$
 $\forall i,j$, consequently every function is substantial here (cf.
 Fig.~\ref{fig:ChuaEmbRecFuncs}).  Given that we analyzed 50,000 data points,
 the $C_{ij}$ refer to a very high correlation.  Therefore we  infer a
 property of the embedding transformation: each of the embedding space
 coordinates $\xi_i$ contains information necessary for the dynamics.

iii) \underline{\it The nine functions} $\phi_{ij}$, displayed in
Fig.\ref{fig:ChuaEmbRecFuncs} are the most important result for an
application. All functions are important and nonlinear, to a good
approximation of cubic order; only $\phi_{13}$ appears to be a piecewise
linear function.
The quantitative comparison of the dynamics of the reconstructed and the
original system is done by i) calculation of the fixed points, ii) their
stability and iii) the Lyapunov exponents (LE's) of the reconstructed system.
These quantities have to coincide with the ones of the original system.

\underline{\it i) Fixed points.}  We solved $\sum_{j=1}^3 \phi_{ij} =0$
($i=1,2,3$) numerically with the functions from the output of the
analysis. The three fixed points of the embedded system are
$(-7.75,-7.75,-7.75)$, $(0,0,0)$, $(7.75,7.75,7.75)$ with an accuracy of
$10^{-3}$ . In the system, reconstructed with $\tau=0.2$, the fixed points are
$\vec{\xi_1^{\ast}} = (-7.76,-7.55,-7.66)$, 
$\vec{\xi_2^{\ast}} = (7.75,7.52,7.68)$, and 
$\vec{\xi_3^{\ast}} = (0,0,0)$
with an error of less than 1\%.

\underline{\it  ii) Stability analysis.}
The  eigenvalues, corresponding to the above fixed points are
$\vec{\gamma}_1= (-7.68,0.47+i\,4.45, 0.47-i\,4.45)$,
$\vec{\gamma}_2= (-7.62,0.58+i\,4.55, 0.59-i\,4.55)$,
$\vec{\gamma}_3= (5.09,-1.16+i\,4.56,-1.16-i\,4.56)$ 
to be compared with the ones of the {\it original} Chua system:
$\vec{\gamma}_{o,1}=(-8.76,0.28+i\,5.20,0.28-i\,5.20)$,
$\vec{\gamma}_{o,2}=(-8.76,0.28+i\,5.20,0.28-i\,5.20)$,
$\vec{\gamma}_{o,3}=(5.03,-1.21+i\,4.71,-1.21-i\,4.71)$.  
For the embedded attractor, there is nothing to calculate due to missing
equations. Furthermore,  the embedding conserves dynamical properties. The
contraction rate from $\vec{\gamma}_{1,2}$ is found within 15\%, the expansion rate
from $\vec{\gamma}_3$ is found within 1\%, the imaginary parts
coincide within 15 \%. 

\begin{figure}
  \begin{center} 
      \includegraphics[draft=false,clip=true,angle=0,width=0.4\textwidth]
		      {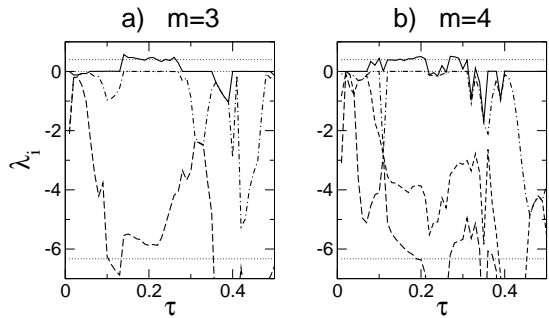}
  \end{center}
  \caption{Lyapunov exponents for original and reconstructed system for
 embedding dimension m=3 (a) and m=4 (b). Increasing $m$ results in a larger
 window in the delay time for which the system is reconstructed, i.e. the LE's
 coincide well. The thin dotted lines indicate the LE's for the original
 system, $\lambda_1=0.432$, $\lambda_2=0$, $\lambda_3=-6.31$, the straight,
 dash-dotted and dashed lines the correspondent ones for the reconstructed
 system.}
\label{fig:ChuaEmbRecLyap}
\end{figure}

\underline{\it iii) Lyapunov exponents and dependence on the delay.}  We
calculated the Lyapunov exponents of the reconstructed system for
$0<\tau\leq 1$. For most of the delays no useful reconstruction is possible,
however in the window $0.14<\tau<0.28$ the LE's are very close to the original
ones (Fig.~\ref{fig:ChuaEmbRecLyap}a).  By eye, it is hard to recognize  which
attractor is reconstructed or embedded (Fig.~\ref{fig:ChuaEmb}).
With this study, we have determined the delay which is optimal in the sense of
nonparametric embedding. Usually, the delay is chosen such that the
information content in the delay coordinate vector is maximized. To do so one
determines the minimum of the mutual information or the first zero of
autocorrelation or similar measures \cite{Kantz-Schreiber-97}.  It turns out
that these approaches do not yield a delay different from ours.

If the embedding dimension is increased, one expects a good reconstruction in
a larger delay-time window, because more information is used.  This is
confirmed by the calculation of the LE's with $m=4$,
(Fig.~\ref{fig:ChuaEmbRecLyap}b), where a good reconstruction is found for
$0.08<\tau<0.36$. The attractor for $\tau=0.2$ is shown for
comparison(Fig.~\ref{fig:ChuaEmb}, top).

At $m=4$, there is a breakdown of the reconstruction for
$0.22<\tau<0.26$, whereas $m=3$ yields good results
(Fig.~\ref{fig:ChuaEmbRecLyap}). This is unexpected and  a
conclusive explanation requires further investigation.

\begin{figure}
  \begin{center} 
      \includegraphics[draft=false,clip=true,angle=0,width=0.4\textwidth]
		      {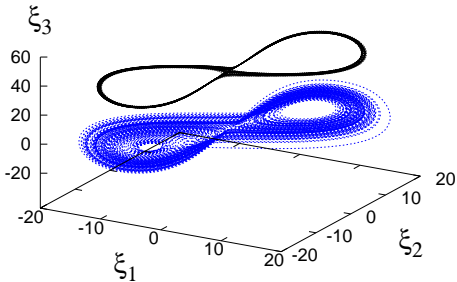}
  \end{center}
  \caption{Reconstruction for the Lorenz system ($m=3$). A part of the
  dynamics is not reconstructed, but may be recovered with higher dimensional
  embedding.}
\label{fig:lorenz}
\end{figure}

A particularity of the modified Chua system is its additive structure.  Next,
we check whether a successful reconstruction can be found for dynamical
systems with multiplicative terms, too, such as the Lorenz or the R\"ossler
system. For both, we find a worse capability of our method to reconstruct the
dynamics. For the Lorenz system ($\sigma = 10$, $\rho = 28$, $\beta =
8/3$), one of the best results appears for $\tau=0.09$, the corresponding
reconstructed attractor is shown in Fig.~\ref{fig:lorenz}. Clearly, some of
the dynamics is lost, nevertheless a chaotic motion about the correct fixed
points is found. The largest LE is found to $\lambda_{rec}=0.08$ to be
compared with the original one, $\lambda_{o}=0.905$. Tests with $m=4$, and
$m=5$ did not yield significant improvement. The reason lies probably in the
topology of the attractor which cannot be produced by a purely additive model
of reasonably low dimensionality. This is a limit of the additive modeling
approach.

%
We have successfully reconstructed a dynamical system by a set of ordinary
differential equations in embedding space. We have considered additive models
only, and have used as a typical chaotic system a modified Chua oscillator for
illustration. The resulting equations can be analyzed by dynamical systems
theory: we have investigated the fixed point structure, linear stability, and
Lyapunov exponents and have found that these dynamical characteristics
quantitatively coincide with the ones of the original system. By studying the
dependence of the results on the delay-time, we could identify the window in
which our method works very well. Higher embedding dimensions enlarge this
window, over-determination can, however, let the description break down. For
non-additive systems, our analysis works qualitatively, a quantitative
comparison is in general not possible, although the result indicates which
terms can be important in a more general model.

In the method, the statistical backfitting algorithm is used for an estimation
of the CEV; the result is a set of optimal functions $\phi_{ij}$. It is
inherently insensitive against noise \cite{Backfit,Abel-04} and can be
generalized in many ways. The results are functions of one variable and can be
visualized and approximated by analytical formulae {\it after} the backfitting
procedure. This yields an important advantage: when fitting polynomials or
other basis systems one chooses these functions beforehand, this is not needed
in the nonparametric approach, and the result is still interpretable. From a
practical point of view, the input data are crucial for a good estimation on a
connected region and estimation of derivatives. Asymptotics and gaps due to
missing data have to be treated with great care or instead of derivatives one
might rather use a mapping approach \cite{Lichtenberg-Lieberman-92}. Decision
on additivity of a model works by statistical measures (correlations or
least-squares error), whereas dynamical measures, as Lyapunov Exponents,
indicate how well the dynamics is reproduced.

From a theoretical point of view, we formulate the following, general
question: given a dynamical system (additive/multiplicative structure), which
topology of a corresponding attractor is possible? Vice versa, given a
topology and dynamics, which is the structure of the underlying dynamical
system?  We have treated a given topology (of the Chua, Lorenz and R\"ossler
system); taking into account the embedding theorem, the problem is transferred
to embedding space. There, we have reconstructed a dynamical system of additive
structure, successfully for the Chua system, less convincingly for the Lorenz
and R\"ossler system. This suggests that the additive structure is kept. Mathematically,
related questions have been treated in \cite{KV}. A key role is played by the
nonlinear embedding transformation which can distort the system
considerably. To our knowledge the above questions are open and touch the core
of modern theory of dynamical systems.

Current and future activities focus on generalization to reconstruct mixed
additive/multiplicative models following Kolmogorovs ideas, especially for
real data.  One goal is to follow the way from the general model
(\ref{eq:embflow}) to a purely additive model (\ref{eq:AddModel}). Finding the
model which involves the least possible multiplicative and additive terms
yields a considerable ease to analyze the systems. With an analytical
expression a detailed analysis and long-term prediction of a (chaotic) orbit
is possible, this is an unprecedented result. Applications for our method
reach from geophysics and climatology to biology and medicine, where the
prediction of, e.g., climate change or illness detection are topics of
superior interest.

We thank B. Fiedler, P. Grassberger, M. Holschneider, and A. Pikovsky for
helpful discussion. M. A. and K. A. are supported by the German science
foundation, DFG; J. K and S. M.   acknowledge a EU grant (HPRN-CT-2000-00158).
                                 
\bibliographystyle{apsrev}

\end{document}